\documentclass[aps,pre,twocolumn,groupedaddress,showpacs]{revtex4}
\usepackage{amsmath,amssymb}
\usepackage[dvips]{graphicx}
\newcommand{\beqar}{\begin{eqnarray}}
\newcommand{\eeqar}{\end{eqnarray}}
\newcommand{\beq}{\begin{equation}}
\newcommand{\eeq}{\end{equation}}
\newcommand{\ket}[1]{\left|{#1}\right\rangle}
\newcommand{\bra}[1]{\left\langle{#1}\right|}

\newcommand{\aver}[1]{\left\langle{#1}\right\rangle}
\newcommand{\ad}{a^{\dagger}}
\newcommand{\bad}{b^{\dagger}}
\newcommand{\added}[2]{\left|{#1},{#2}\right\rangle}
\newcommand{\inner}[2]{\left\langle{#1}|{#2}\right\rangle}
\begin{document}
\title{Recurrence properties of quantum observables in wave packet 
dynamics}

\author{C. Sudheesh}
\email{sudheesh@math.iisc.ernet.in}
\affiliation{Department of Mathematics, Indian Institute of Science,
Bangalore 560 012,  India}
\author{S. Lakshmibala}
\email{slbala@physics.iitm.ac.in}
\author{V. Balakrishnan}
\email{vbalki@physics.iitm.ac.in}
\affiliation{Department of Physics, Indian Institute of Technology Madras,
Chennai 600 036, India}
\date{\today}
\begin{abstract}
We investigate the recurrence properties  
of the time series of quantum mechanical expectation values,  
in terms of two representative models for
a single-mode radiation  field interacting with a nonlinear
medium. From recurrence-time distributions, return maps and recurrence 
plots, we conclude that the dynamics of appropriate 
observables pertaining to the field 
can vary from quasiperiodicity to hyperbolicity, depending on the extent
of the nonlinearity and of the departure from coherence of the initial state of 
the field.
We establish that, in a simple bipartite model 
in which the field is effectively 
an open quantum system, 
a decaying exponential 
recurrence-time distribution, characteristic of a hyperbolic 
dynamical system, is associated
with chaotic temporal evolution as characterized by a positive
Liapunov exponent. 
\end{abstract}

\maketitle

\section{Introduction}
\label{introdn}

Poincar\'e recurrences of 
classical dynamical systems yield a great deal of   
information on their ergodicity properties\cite{kac}. 
Recurrence-time statistics 
complements and augments the information obtained from 
other well-known signatures and 
quantifiers 
of dynamical behavior  such as 
return maps, Liapunov spectra, time-series 
analysis, generalized dimensions, and so on. 

There exists a body of  
rigorous results in the theory of dynamical 
systems that pertains to universal properties of 
recurrences. Most notably, 
for Axiom-A systems\cite{sina,coll,hira1} 
and more generally for uniformly hyperbolic 
systems\cite{hira2}, 
the recurrence time to a sufficiently 
small cell in phase space is exponentially 
distributed. Moreover, successive recurrence times are 
independently distributed, so that the sequence 
of these recurrence times has a 
Poisson limit law. These 
are familiar features of stochastic systems
(e. g., aperiodic 
Markov chains\cite{fell,pits,cox}), but the 
significant  point is that they  
are also exhibited by deterministic 
systems with a sufficient degree of mixing. 
More detailed studies of recurrence-time 
statistics have been carried out in the 
case of low-dimensional chaotic systems, in particular, 
in the framework of one-dimensional maps. 
These investigations enable us to distinguish clearly between 
different degrees of randomness in the dynamics, ranging 
from quasiperiodicity through intermittency to 
fully-developed chaos\cite{vb1,cnico}. Departures 
from the standard results for hyperbolic systems 
have been analyzed, showing (for instance) that 
intermittency leads to power-law  
(as opposed to exponential) distributions of 
recurrence times, with correlations between 
successive recurrences\cite{vb2} and non-Poisson 
limit laws\cite{vb3}.  A variety of results is also known 
for recurrence-time distributions for 
Hamiltonian systems and for measure-preserving maps 
that model aspects of such systems. Interesting universal 
asymptotic properties, including power-law distributions 
of recurrence times, can arise here owing to the 
highly non-uniform nature of the invariant sets in 
phase space and the stickiness associated with 
the remnants of invariant 
tori\cite{afrai,chiri,zasla,buric1,buric2,altmann1,altmann2,artuso}. 

Besides recurrence-time distributions, 
recurrence plots comprise a related technique 
for analyzing dynamics ranging from periodicity 
to chaos\cite{eckm,marwan07}. Several distinctive features 
of recurrence plots have been identified and established as 
indicators of specific kinds of dynamical behavior such as 
multiple periodicity, intermittency, and chaos. 
Recurrence quantification analysis\cite{marwan02} has 
been developed in an attempt to deduce quantitative 
information from the heuristics of recurrence plots. 

In contrast to these detailed results 
for classical dynamical systems,  
the ergodicity properties of  
expectation values of observables 
in nonlinear quantum systems, regarded as dynamical 
variables,  
are much less comprehensively understood.  
A fundamental difficulty  that immediately arises here is that 
the `phase space' is effectively infinite-dimensional, 
because the complete information
contained in a
specific quantum state can be obtained, 
in principle,  only 
if the mean values as well as 
all the higher moments and cross-correlators of all the 
quantum operators pertaining to the
system are included in the set of  dynamical variables. 
Moreover, significant roles are played 
by the precise  
nature  of 
the initial state of the system,  
and by the interaction of the system of 
interest with the environment. 
Owing to these factors (and, of course, the 
non-commutativity of different operators),  
the  dynamics of 
expectation values may be expected to 
be quite complex, even in relatively  
simple systems.  It is therefore relevant to 
examine specific tractable models of 
quantum dynamics in order to 
disentangle the effects of various factors, and to 
discern systematic trends.

The time evolution of quantum mechanical 
wave packets  provides an 
appropriate framework for 
the purposes described above. 
The phenomenon of revivals\cite{robi} 
provides a manifest analog of recurrences in coarse-grained 
dynamical systems. An initial state 
$\ket{\psi(0)}$ has a revival time 
$T_{\rm rev}$ 
if the overlap function
 $|\inner{\psi(0)}{\psi(T_{\rm rev})}|^2$ 
 returns to an $\epsilon$-neighborhood 
of its initial value of unity. 
A physical realization of this phenomenon is provided by the 
propagation of a radiation field in a nonlinear medium 
under suitable conditions. 
A wave packet of the radiation 
spreads 
and loses its original form almost immediately owing to 
the field-atom interaction, but displays revivals at 
integer multiples of 
$T_{\rm rev}$. Correspondingly, all expectation 
values return to the neighborhoods of their initial values. 
Distinctive signatures of the revivals, and of 
the frational revivals in   between successive 
revivals, can be identified by analyzing  the 
(projections of the) `phase trajectories' of the system 
in subspaces of appropriate expectation values\cite{sudh2004}.
Typically, the behavior of these trajectories is similar to 
that of the phase trajectories in quasiperiodic 
dynamical systems with 
many  incommensurate frequencies. 

However, departures 
from this scenario occur with increasing 
deviation of the initial state from perfect coherence, or increasing
nonlinearity of the medium, or both: It is 
found\cite{sudh5} that  regular revivals of 
wave packets no longer occur under these conditions, namely, 
high nonlinearity and non-coherent initial states. A 
more detailed examination of this physically 
interesting, and obviously  more generic, regime is   
called for. This is the primary objective of the present paper.  
The objects we study are the time series formed by the expectation
values of relevant operators. The aspects we focus on are the
recurrence-time  or first-return-time 
statistics of these time series, and
their recurrence plots. 

In what follows, we use two simple models 
that capture the essence of 
the effects we seek to clarify and highlight. 
In the first of these,  
considered in Sec. \ref{singlemodemodel},  
we use the fact\cite{tanas} that certain aspects of 
the dynamics of the radiation field propagating in a nonlinear medium can 
be modeled by an effective Hamiltonian involving the field operators 
alone, without explicitly invoking the atomic operators. 
The purpose here is to understand the effects of the lack of 
coherence of the initial state of the 
wave packet upon its subsequent evolution and upon 
recurrence-time distributions.  The Hamiltonian concerned has 
a purely discrete spectrum, which 
implies, according to a rigorous result\cite{kosloff},  
that the dynamics of expectation values cannot
be chaotic (the maximal Liapunov exponent cannot be positive). 
Nevertheless, we find that, 
with increasing nonlinearity 
and/or lack of coherence in the initial state, 
the recurrence-time distribution  of a generic 
 observable changes from 
a form that is characteristic of 
quasiperiodicity to an exponential distribution, 
which is customarily associated with 
hyperbolic (and hence chaotic) motion in 
classical dynamics. An exponential recurrence-time 
distribution is  therefore not restricted to 
chaotic systems.  

While this model serves to illustrate the point 
just made,  
it must however be recognized  that
the radiation field is really one 
of the two components  of a bipartite,  
interacting,  field-atom system. In order to 
investigate what happens in this situation,   
we consider, in Sec. \ref{twomodemodel},  
a   bipartite Hamiltonian that involves 
both the field 
operators and  the atomic ladder operators.  
Focusing on 
operators involving the field mode alone then 
amounts, effectively,   to 
considering an {\it open} quantum system interacting 
with the (atomic) environment.  Even if
the initial state is a non-entangled  
direct product state
of the two modes, 
entanglement  sets in
during temporal evolution.
The controlling parameter in this case 
is the ratio of the 
respective strengths of the nonlinearity and the 
inter-mode coupling.  For 
low values of this ratio, an  
entangled state periodically factorizes
into the
initial product state (apart from an overall phase), 
and revivals occur. 
For higher values of the ratio, 
such  revivals disappear. Depending on  
how far the initial state of the field is from perfect coherence,  
the dynamics of relevant expectation
values emulates that of a hyperbolic dynamical system. 
Recurrence-time distributions, recurrence plots, as 
well as a calculation of the maximal Liapunov exponent 
based on phase-space reconstruction using the 
time series of an expectation value, all corroborate this 
conclusion, which therefore has implications for the 
dynamical behavior of open quantum systems.

\section{Single-mode nonlinear Hamiltonian}
\label{singlemodemodel}
    
As stated in the Introduction, we begin with 
a simple effective Hamiltonian for a 
single-mode electromagnetic 
field 
interacting with the atoms of a
nonlinear medium, 
\begin{equation}
H = \hbar\,( \chi \,\ad{^2} a^2 + \chi\,' \,\ad{^3} a^3), 
\label{hamiltonian1}
\end{equation}
where the photon  annihilation and 
creation operators satisfy $[a, \ad] = 1$.
The first term in $H$ is the 
standard one modeling a Kerr medium, with a coupling strength $\chi$. 
The second term is an additional 
cubic nonlinearity with a strength $\chi\,'$. 
It introduces a second natural 
frequency into the system. Since $\chi\,'/\chi$ is arbitrary and
generically not a rational number, it  
modifies the relatively simple behavior in the absence of 
such a term. Both terms are of course diagonal in the number 
operator $\ad a$, since 
$\ad{^2} a^2  = \ad a(\ad a-1)$ and 
$\ad{^3} a^3 = \ad a(\ad a -1)(\ad a-2)$.  As 
a consequence, the mean photon number in any state of the 
system remains equal to its initial value as the state 
evolves in time. 

We first 
recapitulate very briefly how revivals occur in the 
absence of the cubic nonlinearity,  i.e., when  
$\chi\,'= 0$. At the instants $t = \pi/(k \chi)$ (where $k$ is 
a positive integer), the time-evolution operator $U(t) = 
\exp\,(-iH t/\hbar)$, which is diagonal in the Fock basis, has matrix 
elements $\exp\,[-i\pi n(n-1)/k]$, where $n$ denotes the 
eigenvalues of $\ad a$. Even though the exponent 
here is quadratic rather than 
linear in $n$, this exponential   has 
interesting periodicity properties\cite {aver} 
as a function of $n$.   
As a consequence of these properties, 
$U(\pi/k \chi)$ can be expanded in a  Fourier
series with $\exp\,(-2 \pi i j/k)$ as the basis functions, where 
the integer $j$ runs over the range 
$1 \leq j \leq (k-1)$.  If the initial wave packet is 
the Gaussian 
wave packet corresponding to a 
minimum-uncertainty coherent state (CS) 
$\ket{\alpha}$ (where 
$a\ket{\alpha} = \alpha \ket{\alpha}$ and 
$\alpha \in \mathbb{C}$), then 
the state at time $\pi/(k\chi)$ 
can be written as a finite linear combination of 
the states $\ket{\alpha\,e^{-2\pi i j/k}}$ (for odd $k$)  
or the states $\ket{\alpha\,e^{i\pi/k}\,e^{-2\pi i j/k}}$ 
(for even $k$). But these are again coherent states, 
with parameters that are just phase-shifted versions of the 
original parameter $\alpha$. It is then easy to 
see that, at  time $T_{\rm rev} = \pi/\chi$
(corresponding to $k = 1$), 
the initial state revives for the first time. Periodic
revivals occur at integer multiples of $T_{\rm rev}$.
 Fractional revivals are also evinced at 
 specific 
 times in between successive revivals. 
 
The time evolution is quite different when 
$\chi\,' \neq 0$. For generic initial wave packets and 
(irrational) values of the ratio  $\chi\,'/\chi$, 
exact revivals do not occur. 
Correspondingly, in the 
space of observables, 
periodic returns of 
observables to their 
initial values is replaced by   
quasiperiodicity. Now, the 
dynamical behavior of expectation values in this model 
cannot become chaotic for any parameter values 
(as already mentioned), because 
the evolution is 
governed by a quantum Hamiltonian with a purely 
discrete  spectrum\cite{kosloff}.   
However, as the 
mean photon number increases, or the initial state 
departs from perfect coherence, the quasiperiodicity 
involves an increasing number of 
incommensurate frequencies. Concomitantly, the 
recurrence-time distributions and  recurrence  plots  
go over from the forms characteristic of quasiperiodity 
to those pertaining to a dynamical system with a 
much higher degree of mixing, as we shall see below.  

Our aim is to examine the role played by the extent of 
coherence of the initial state on the recurrence properties. 
An obvious dynamical variable for this purpose is the 
expectation value of the 
quadrature $x = (a + \ad)/\sqrt{2}$.
The  standard  
CS $\ket{\alpha}$  serves as the 
reference initial state of the field mode. 
The mean photon number in this state is 
of course $|\alpha|^2$, which we shall denote by $\nu$. 
The non-coherent initial states
we consider are 
photon-added coherent states (PACS)\cite{tara}.  
 The normalized $m$-photon-added coherent state
 $\added{\alpha}{m}$ 
 has  a quantifiable 
 (and, in principle, tunable)  
departure from perfect coherence and 
Poissonian photon statistics.  It is defined as
$\added{\alpha}{m}
=(a^\dagger)^m\ket{\alpha}/[m!\,L_{m}(-\nu)]^{1/2}$, 
where $m$ is a positive integer 
and $L_{m}$ is
the Laguerre polynomial
 of order $m$.  The mean photon number in this 
 state is  given by $(m+1)[L_{m+1}(-\nu)/L_m(-\nu)] - 1$.  
 Moreover, the variance of the photon number 
 can be shown to increase sublinearly 
with the mean photon number, 
 implying that the photon statistics is sub-Poissonian.  
 With the
experimental production and
characterization of a 1-photon-added
coherent state by quantum state
tomography\cite{zavatta},  
photon-added coherent
states are expected to  play an increasingly
 significant role in future
investigations.

\begin{figure} 
\includegraphics[width=3.0in]{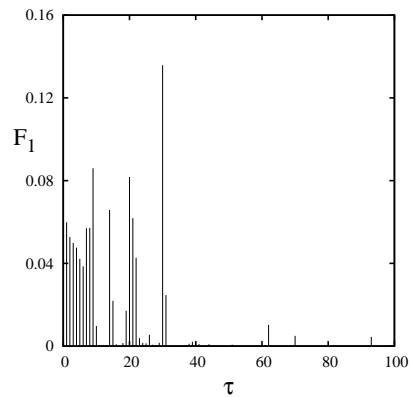}
\caption{\label{fig1} First return time distribution $F_1$ of 
the expectation value $\aver{x}$. Initial  
state 
$\ket{\alpha}$, with 
$\nu  = 1$.}
\end{figure}

\begin{figure}
\includegraphics[width=3.0in]{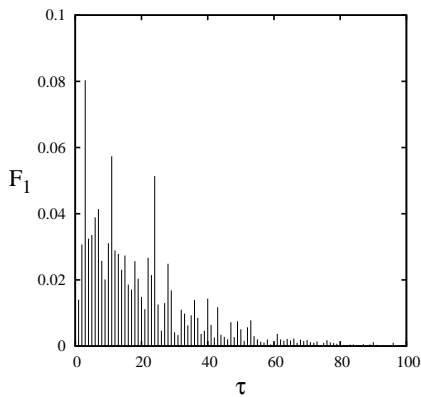}
\caption{\label{fig2}$F_1$ for the initial 
state
$\ket{\alpha,5}$ with 
$\nu  = 1$.}
\end{figure}

\begin{figure}
\includegraphics[width=3.0in]{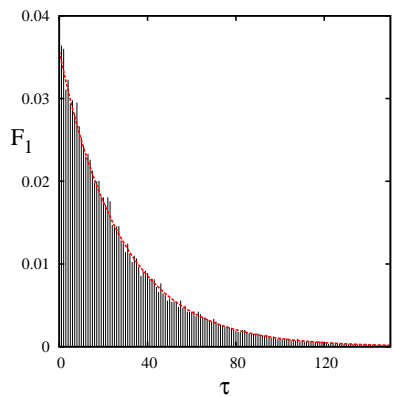}
\caption{\label{fig3}$F_1$ for the initial
state
$\ket{\alpha}$ with 
$\nu = 100$.}
\end{figure}

\begin{figure}
\includegraphics[width=3.0in]{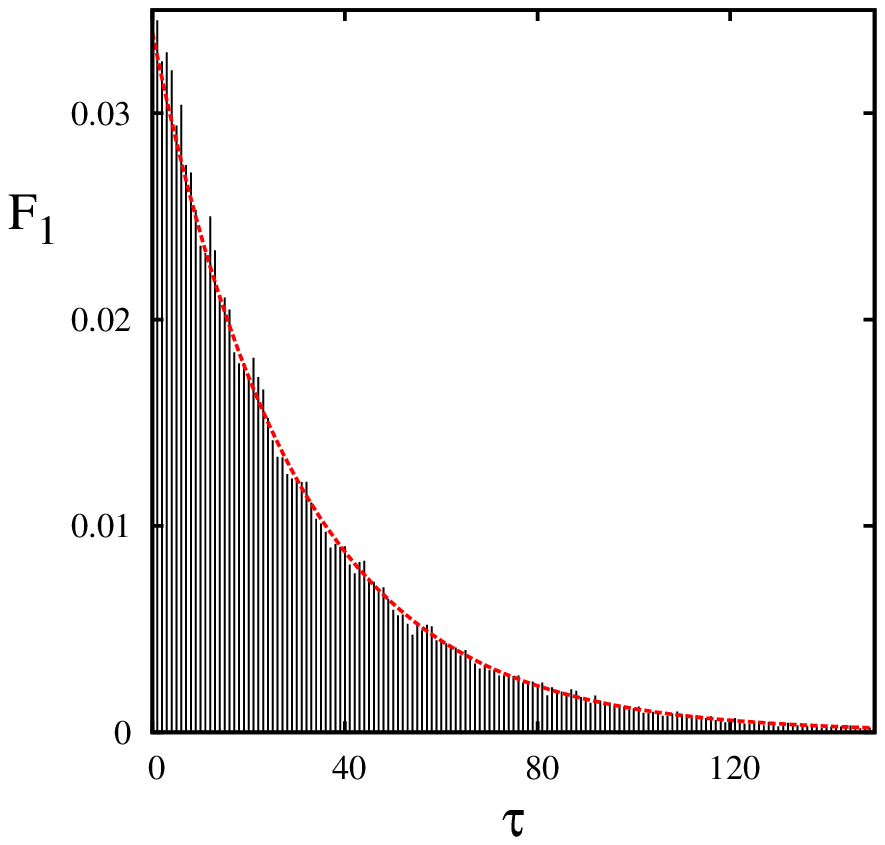}
\caption{\label{fig4}
$F_1$ for the nitial 
state
$\ket{\alpha,5}$ with
$\nu = 100$.}
\end{figure}

\begin{figure}
\includegraphics[width=2.5in]{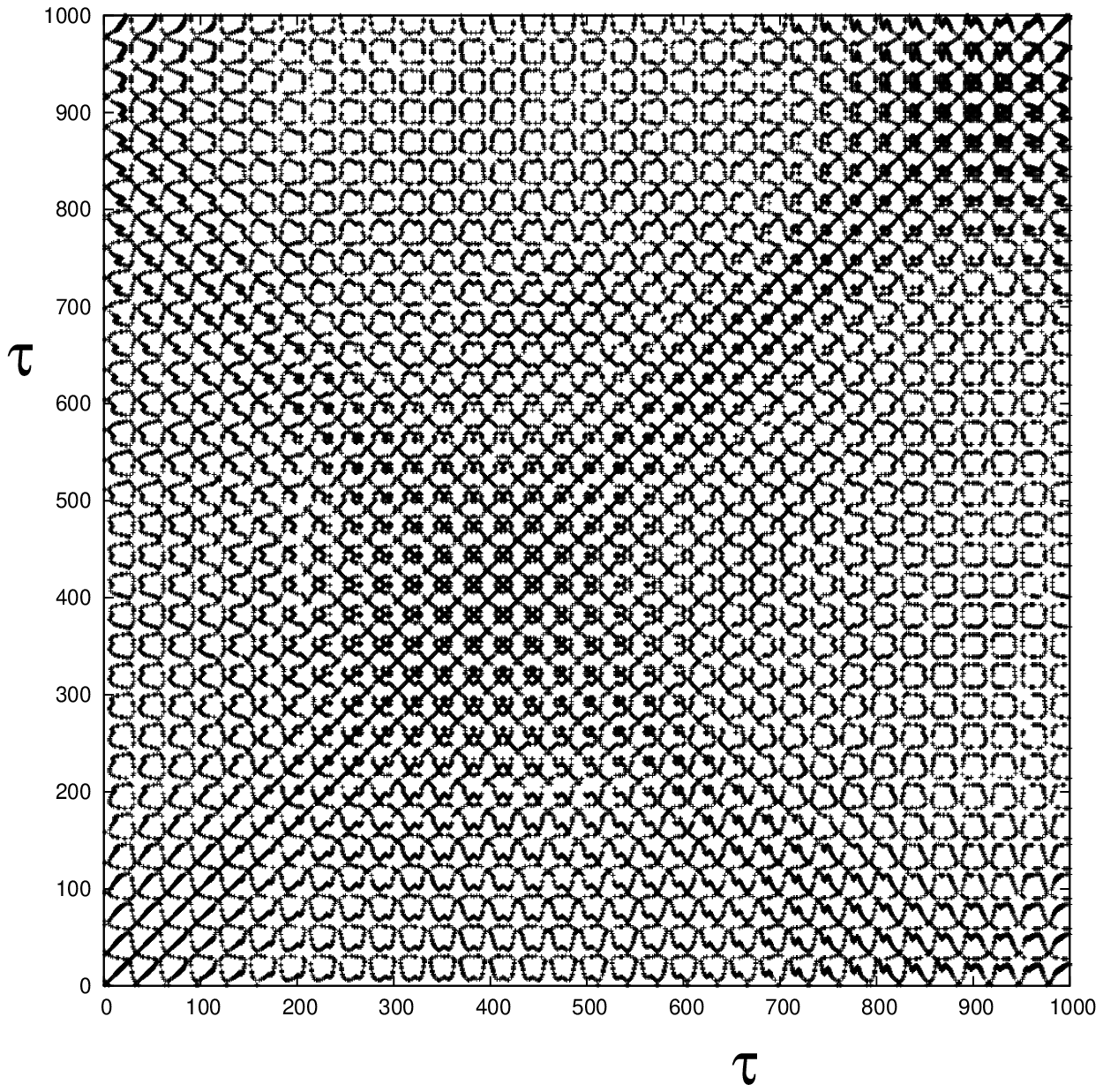}
\caption{\label{fig5} Recurrence plot of the time 
series $\aver{x}$.  Initial state 
$\ket{\alpha}$ with $\nu = 1$.}
\end{figure}

\begin{figure}
\includegraphics[width=2.5in]{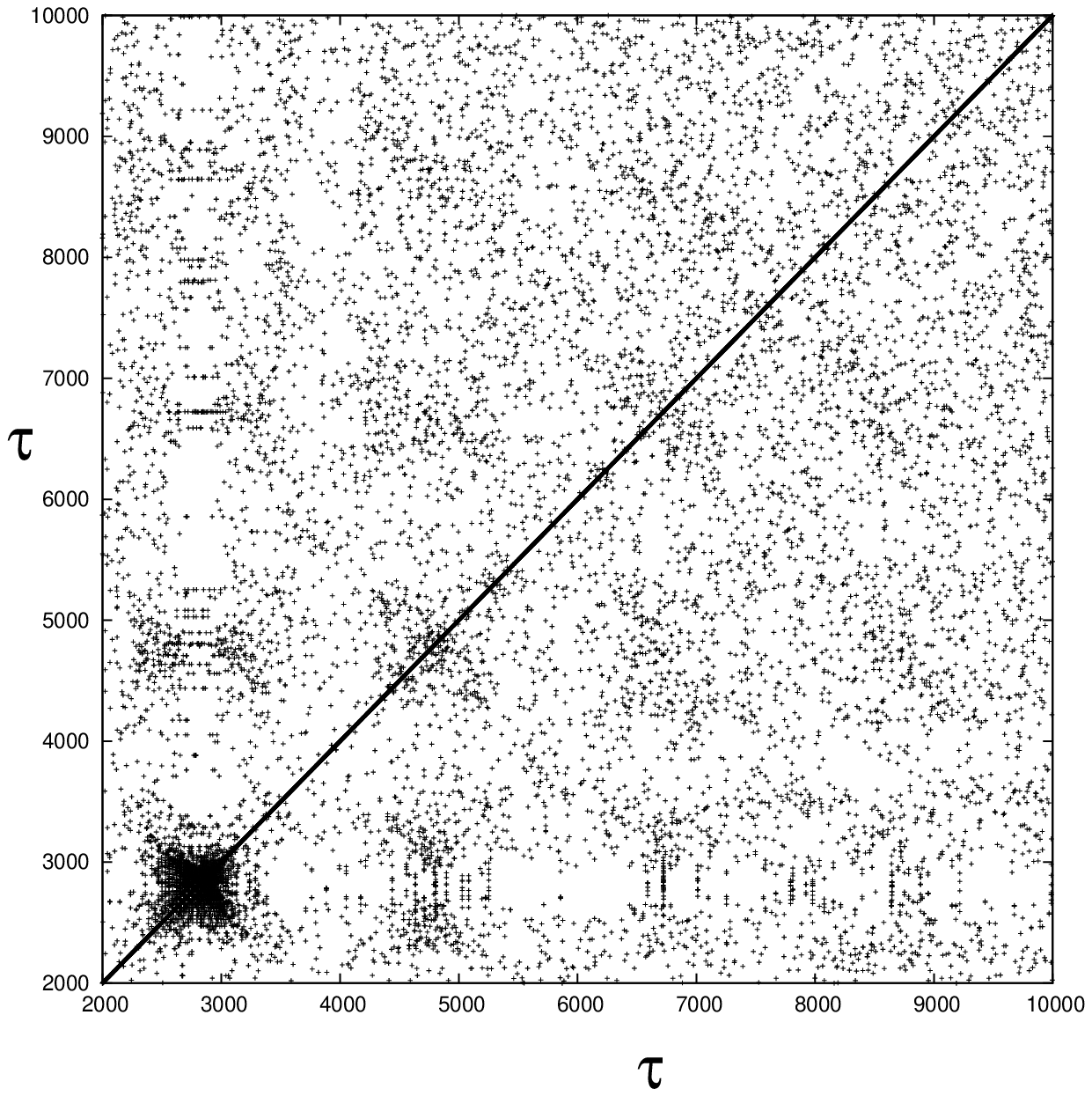}
\caption{\label{fig6} 
Recurrence plot of the time series of 
$\aver{x}$.  Initial state
$\ket{\alpha,5}$ with $\nu = 100$.}
\end{figure}

The procedure we adopt is straighforward. 
We start with a 
given normalized initial state $\ket{\psi (0)}$, let 
it evolve unitarily according to the 
Hamiltonian $H$, and calculate $\langle \psi (t)|x
\ket{\psi(t)} \equiv \aver{x(t)}$ in time steps 
$\delta t$. This produces a 
time series. The interval of values of 
$\aver{x(t)}$ is coarse-grained into 
small cells, and a recurrence  
to a given cell is marked. This enables us to build up 
the recurrence-time distribution for the cell concerned. 
The time series is also used to draw a recurrence plot. 
We have checked that the 
results reported below hold good 
qualitatively for generic cell sizes and 
locations in the phase space considered. For definiteness, 
we present in  Figs. \ref{fig1}--\ref{fig6}  
the  results for a cell size $\delta x = 10^{-2}$ and 
a parameter ratio $\chi\,'/{\chi} = 10^{-2}$. The data 
set comprises a long time series of 
about $10^7$ points,  and the time 
step $\delta t =
10^{-3}$. Time in units of $\delta t$ is 
denoted by $\tau$.
 
We now recall that, for uniform quasiperiodic motion with  
two incommensurate frequencies, the distribution of the 
time of recurrence to any cell (on the $2$-torus) has a 
support comprising just three values\cite{cnico,sesh}. 
This result is derived from a `gap theorem' for 
interval exchange transformations (specifically, 
the rotation map on a circle), and follows from certain  
number-theoretic properties of continued 
fractions\cite{slater,rauzy}. 
For quasiperiodic motion with more than two 
mutually incommensurate frequencies (i. e., on an $n$-torus 
with $n > 2$), the recurrence-time distribution remains a 
discrete distribution with support at a finite number 
of values (although no generally applicable  formula is known for the 
actual number of such values, for arbitrary $n$). 
Returning to the problem at hand, 
we find that when the initial state is a CS $\ket{\alpha}$, 
with a mean photon number $\nu$ that is of the order of unity, 
the recurrence-time  (or first return time) distribution $F_1$ of 
$\aver{x}$ is a discrete distribution with support at a small, finite 
number of points, as in Fig. \ref{fig1}. 
If, instead of the CS $\ket{\alpha}$, 
we use the PACS  
$\added{\alpha}{m}$    
as the initial state, the points of support of 
the distribution $F_1$ increase  in number  
with increasing $m$. Figure \ref{fig2} illustrates the case 
$m = 5$. 
As $m$ takes on larger values, 
this trend rapidly leads to an essentially continuous, 
decaying exponential distribution. The same trend occurs even for an
initially coherent state $\ket{\alpha}$, provided the mean 
photon number $\nu \gg 1$, as seen in Fig. \ref{fig3}. The onset 
of an exponential distribution is even more pronounced for 
an initial PACS, as may be seen  in Fig. \ref{fig4}. 

The recurrence plots obtained from the time series 
for $\aver{x}$ are completely consistent with this 
scenario. Figure \ref{fig5} depicts the recurrence 
plot in the case of an
initial CS $\ket{\alpha}$ with $\nu = 1$. The regular, patterned 
structure of the plot is characteristic\cite{marwan07}  
of quasiperiodicity with a relatively small number of 
incommensurate frequencies. In marked  
contrast, Fig. \ref{fig6} shows the recurrence plot for 
an initial PACS $\ket{\alpha,5}$ with $\nu = 100$. 
It is evident that, at the very least, a significant 
degree of  
mixing has set in, corroborating what we have  
already deduced on the basis of the recurrence-time 
distributions. 

We are thus led to the conclusion that the generic dynamics 
in this model makes a gradual transition from 
quasiperiodicity to 
mixing as the mean photon number 
increases, {\em or} if the initial state deviates 
significantly from perfect coherence, 
or both. It has already been mentioned that  
hyperbolically unstable  
dynamics is precluded in this model. We have 
checked this out   
independently by augmenting the foregoing 
with an analysis of the time series
obtained  for $\aver{x}$ along the lines 
customary\cite{grass,fraser,abar}     
in the study of dynamical systems, namely: phase space 
reconstruction, the estimation of the minimum embedding  
dimension of the effective phase space, and the calculation  
of the maximal Liapunov exponent $\lambda_{\rm max}$. 
The result, as expected, is that 
$\lambda_{\rm max} = 0$ to within numerical accuracy. 
The occurrence of an exponential  recurrence-time distribution 
is not inconsistent with this result. As is familiar 
from results for  Markov processes, 
exponential recurrence-time distributions and 
statistically  independent  successive recurrences with 
a Poisson law can occur even in stable stochastic 
dynamics\cite{vb4} with a (mean) maximal Liapunov 
exponent\cite{larnold} 
that is zero or negative. The essential 
requirement is an adequate degree of mixing.

\section{Two-mode model: open quantum system}
\label{twomodemodel}

Having seen how the introduction of the cubic nonlinearity 
in the single-mode model of Sec. \ref{singlemodemodel} 
alters its dynamical behavior, 
we now turn to a more representative model of the 
radiation field propagating in a nonlinear medium. 
This model involves both the field and atom modes explicitly, 
and is given by the bipartite Hamiltonian\cite{puri}  
\begin{equation}
H = \hbar [\omega \,\ad a +  \omega_0 \bad b + 
\gamma \bad{^2} \,b^2 + g (\ad b + \bad a)]. 
\label{twomodehamiltonian}
\end{equation}
As before, $a$ and $\ad$ are the annihilation and creation 
operators for the field. The medium is modeled by an anharmonic 
oscillator with ladder operators $b$ and $\bad$, 
the parameter $\gamma$ serving as a measure of 
the  nonlinearity.  We focus on the relative
roles of the nonlinearity 
and the coupling of the two modes
(quantified by the coupling constant $g$) 
in determining the 
dynamics of observables. In the numerical results 
presented below, we have therefore set  
the unperturbed 
frequencies $\omega$ and  $\omega_0$ equal to unity.  
The relevant tunable parameter is then the ratio $\gamma/g$. 
The idea is to examine the expectation value of an 
operator pertaining to the field alone, which may be 
regarded as an open quantum system in interaction 
with the `environment' represented by the medium. 
Since an open subsystem of a Hamiltonian system is not 
conservative by itself, its evolution is effectively 
guided by a positive map, and any hyperbolicity in the  
dynamics could also lead to chaotic behavior of the 
observable concerned. 

The total
number operator
$ (\ad  a
+ \bad  b)$ commutes
with the Hamiltonian (\ref{twomodehamiltonian})
for all values of the parameters in $H$, but 
the photon number operator $\ad a$
does not do so for any $g\neq 0$.
Thus, while $H$ can be cast in  block-diagonal form in a
direct-product basis of field and atom Fock states, the model is not
trivial.
The Fock basis is given by  
${\ket{n}_{\rm field}\otimes\ket{n'}_{\rm atom}}$,  
where
$n $ and $n'$ are the eigenvalues of $\ad a$ and $\bad b$ 
respectively.
The  basis states can be conveniently taken to be  
$\ket{N-n}_{\rm field} \otimes 
\ket{n}_{\rm atom}
\equiv\ket{N-n; n}$, where 
$N$ labels the eigenvalues of $(\ad a + \bad b)$. 
But $\bra{N-n\,;\, n} H 
\ket{N'-n'\,;\,n'} =
0$
if $N\ne N'$.
Hence, for a given value of $N$, $H$ can be diagonalized in the 
space
of states $\ket{N-n\,;\,n}$,  where $n=0,1,\ldots N$. 
A natural choice for the observable representing the 
field (the `open' subsystem) is the photon number 
operator $\ad a$. 

It is evident that
 if $g=0$, $H$ is just the sum of two decoupled parts. 
If $\gamma=0$, $H$ is linear and hence 
is again trivially 
diagonalizable, and there is periodic exchange 
of energy between the field and  the 
atomic oscillator.
 When $\gamma$ and $g$ are both 
non-zero, the dynamics is more complicated, but the existence
of the conserved 
operator $(\ad a + \bad b)$ ensures that the system as
a whole is well-behaved, with a discrete spectrum labeled
by the quantum numbers $N$ and $s$, where
$N = 0,1,\ldots$ and $s = 0,1,\ldots , N$.
The time-dependence of 
$\aver{\ad a}$ 
is thus a direct consequence 
of the  coupling of the field mode to another degree of 
freedom; and the 
deviation of 
$\aver{\ad a}$ from periodic temporal 
variation is an indicator 
of the effects of the nonlinearity 
in the second degree of freedom.  
For different ranges of values of the 
parameter ratio $\gamma/g$,    
a diversity of temporal 
behavior is exhibited by $\aver{\ad a}$.
In the weakly  nonlinear case $(\gamma/g\ll1)$, collapses 
 and  revivals of 
$\aver{a^{\dagger} a}$ occur almost periodically 
in time for initial field states
which are Fock states or coherent states.
When  $\gamma/g \sim 1$, such collapses 
 and  revivals occur more irregularly
 if the field is initially in a coherent state, 
 as compared to 
an initial Fock state.
In the strongly nonlinear regime $(\gamma/g\gg 1)$, 
collapses and revivals 
do not occur,
and the model 
exhibits more complex dynamical 
behavior\cite{sudh5}.  Our interest here 
is in the  recurrence properties of the 
field observable $\aver{\ad a}$.  
For illustrative convenience,  
we take the atomic oscillator to 
be in the ground state 
and the initial state of the field
to be either a CS or a PACS. We 
shall denote the corresponding states 
of the total system by 
$\ket{\alpha ;0}$
and $\ket{(\alpha, m);0}$, 
respectively. 

\begin{figure}
\includegraphics[width=2.5in]{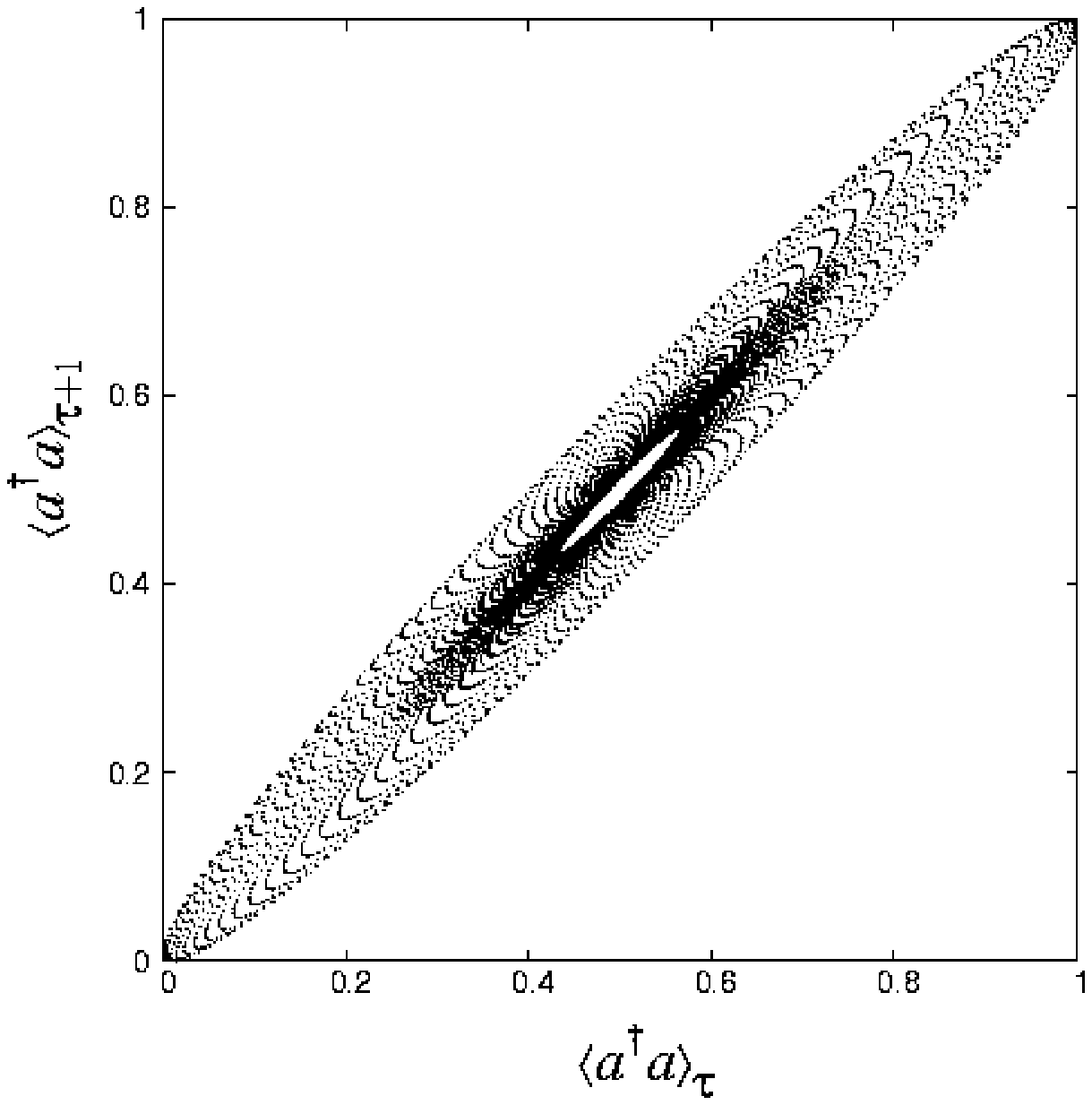}
\caption{\label{fig7} 
Return map of the mean photon number 
$\aver{\ad a}$ for an initial state
$\ket{\alpha;0}$
with  $\nu = 1$;\, $\gamma/g = 10^{-2}$.}
\end{figure}

\begin{figure}
\includegraphics[width=2.5in]{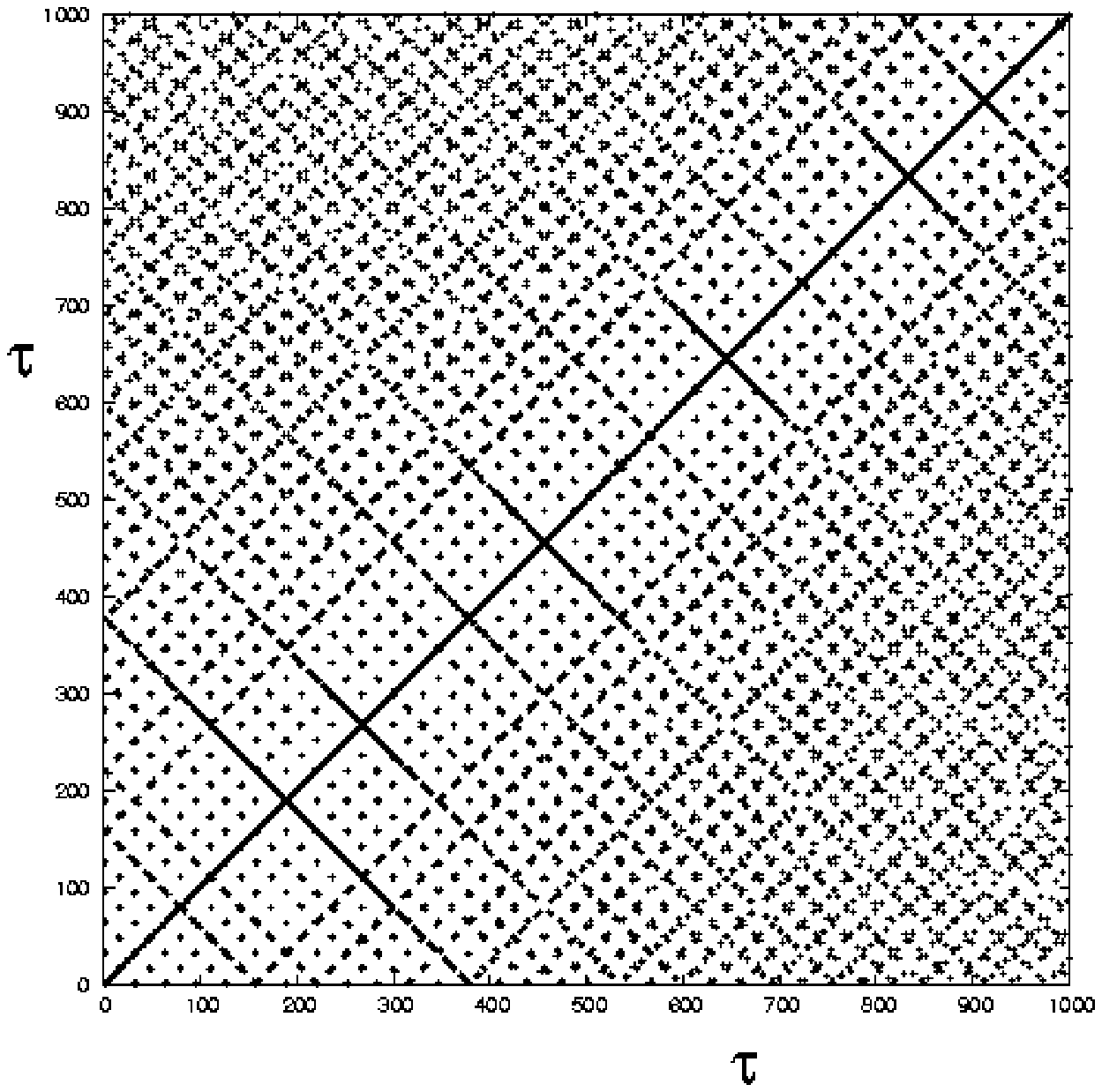}
\caption{\label{fig8} 
Recurrence plot of $\aver{\ad a}$  for the  initial state
$\ket{\alpha;0}$ with $\nu = 1$;\, $\gamma/g = 10^{-2}$.}
\end{figure}

\begin{figure}
\includegraphics[width=2.5in]{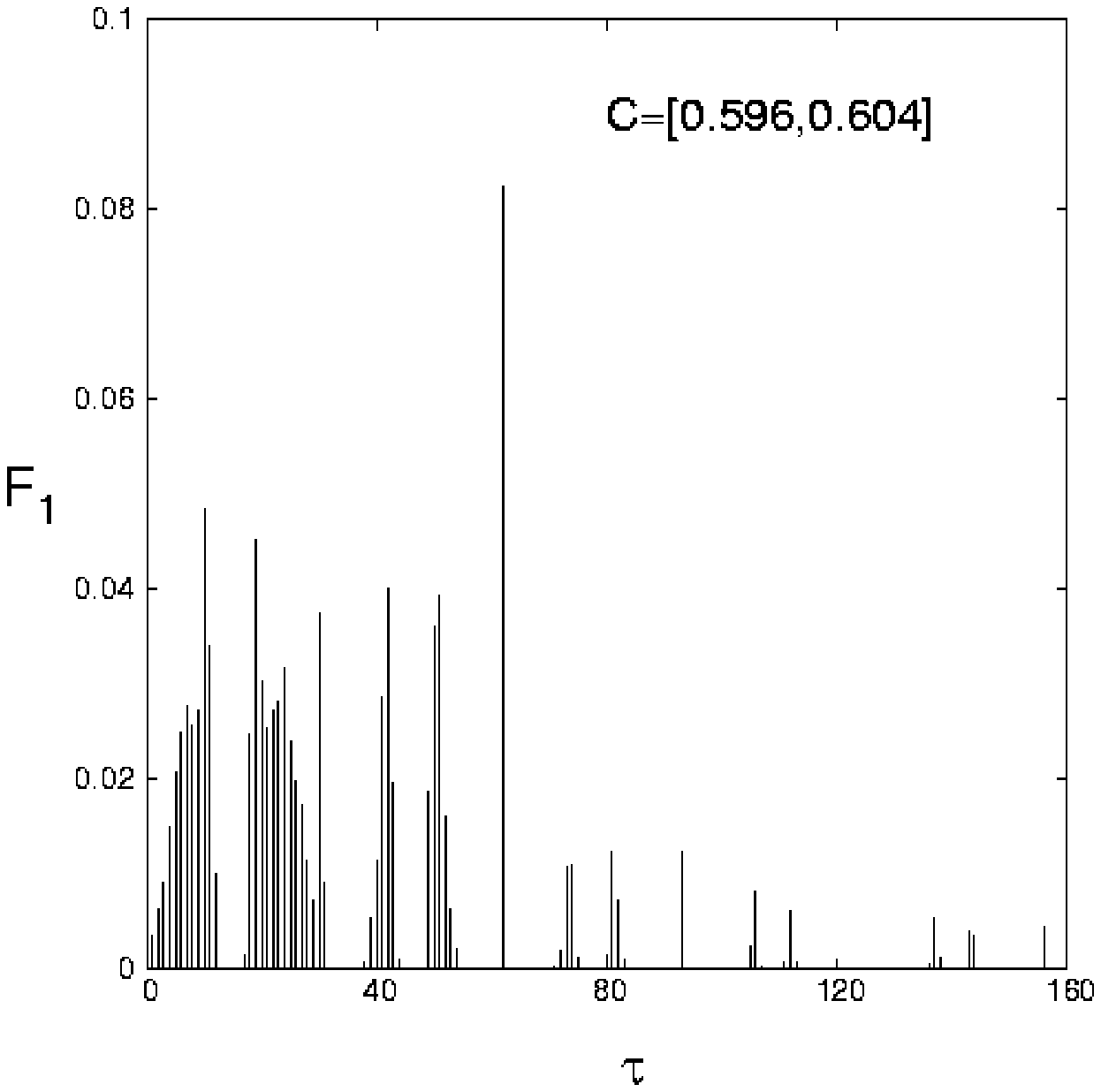}
\caption{\label{fig9} First return time 
distribution $F_1$ of $\aver{\ad a}$ to the cell $C$, 
for the initial state
$\ket{\alpha;0}$ with $\nu = 1$; $\gamma/g = 10^{-2}$.}
\end{figure}

\begin{figure}
\includegraphics[width=2.5in]{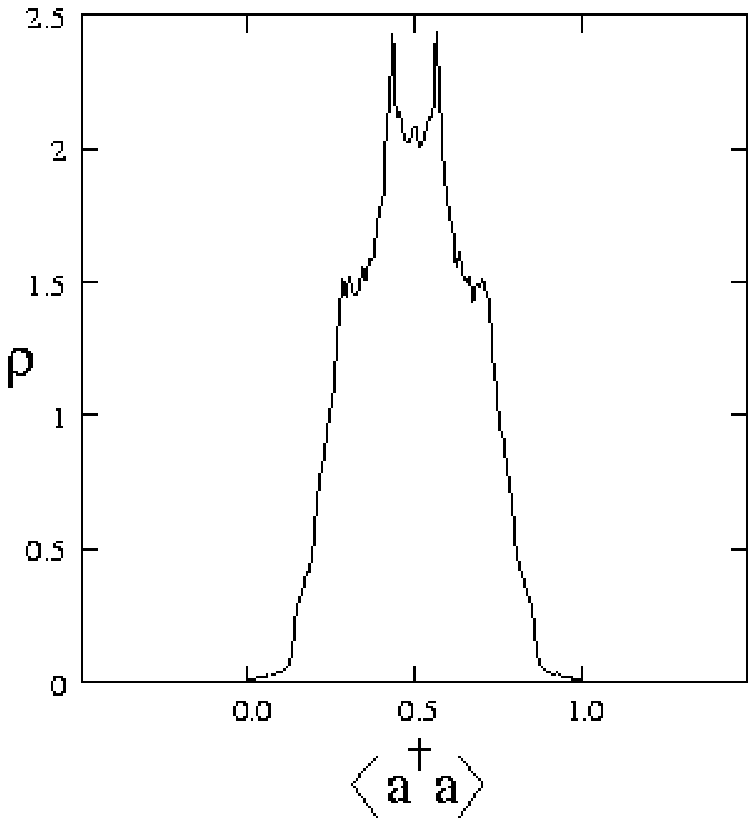}
\caption{\label{fig10} Invariant density 
of $\aver{\ad a}$.  
Initial state 
$\ket{\alpha;0}$ with  $\nu = 1$;\, $\gamma/g = 10^{-2}$.}
\end{figure}

In the case of weak nonlinearity, the return maps, 
recurrence plots, and  first return distributions 
indicate 
quasiperiodicity of the field observables, 
remaining qualitatively independent of whether 
the initial state of the field 
is a CS or a PACS, and of the location and size of the 
cell in the 
coarse-grained `phase space' (i. e., the bin 
in $\aver{\ad a}$).  These results are consistent 
with the occurrence of revivals, and hence 
of returns of observables to their initial values. 
 Figures \ref{fig7}--\ref{fig10}  
correspond to  an 
initial state $\ket{\alpha;0}$, 
with $\nu = 1$ and $\gamma/g = 10^{-2}$. 
The return map 
 in Fig. \ref{fig7} and the 
recurrence plot in Fig. \ref{fig8} reveal regular structures, 
while the distribution of the first return time  
for a typical cell is a sparse set of spikes.  
Figure  \ref{fig9} depicts $F_1$ for the cell $C$ 
defined by  the range $[0.596,  0.604]$ of $\aver{\ad a}$.  
All these features are characteristic of 
quasiperiodic dynamics. A numerical evaluation 
of the (invariant) density 
$\rho$  of $\aver{\ad a}$, the results of which are 
depicted in  Fig. \ref{fig10}, shows that this 
quantity is essentially made up of  a large number of 
spikes. 

For strong nonlinearity ($\gamma/g \gg 1$), 
however, revivals do not occur  for generic conditions,  
and there arises a range  of
ergodicity properties, 
depending on the extent of coherence of the
initial state of the field.  For a significant departure from 
coherence (as measured by the ratio 
$m/\nu$ for an initial PACS), or even for 
 sufficiently large values of $\nu$ in the case 
of an initial CS, the dynamics of $\aver{\ad a}$  
effectively becomes hyperbolic. 
The return map and recurrence time
statistics reveal this feature: the corresponding plots are in
sharp contrast to those that arise in the quasiperiodic case. 
We illustrate this by means of plots for an initial state  
$\ket{(\alpha,5);0}$, with $\nu = 5$ and $\gamma/g =5$. 
\begin{figure}
\includegraphics[width=2.5in]{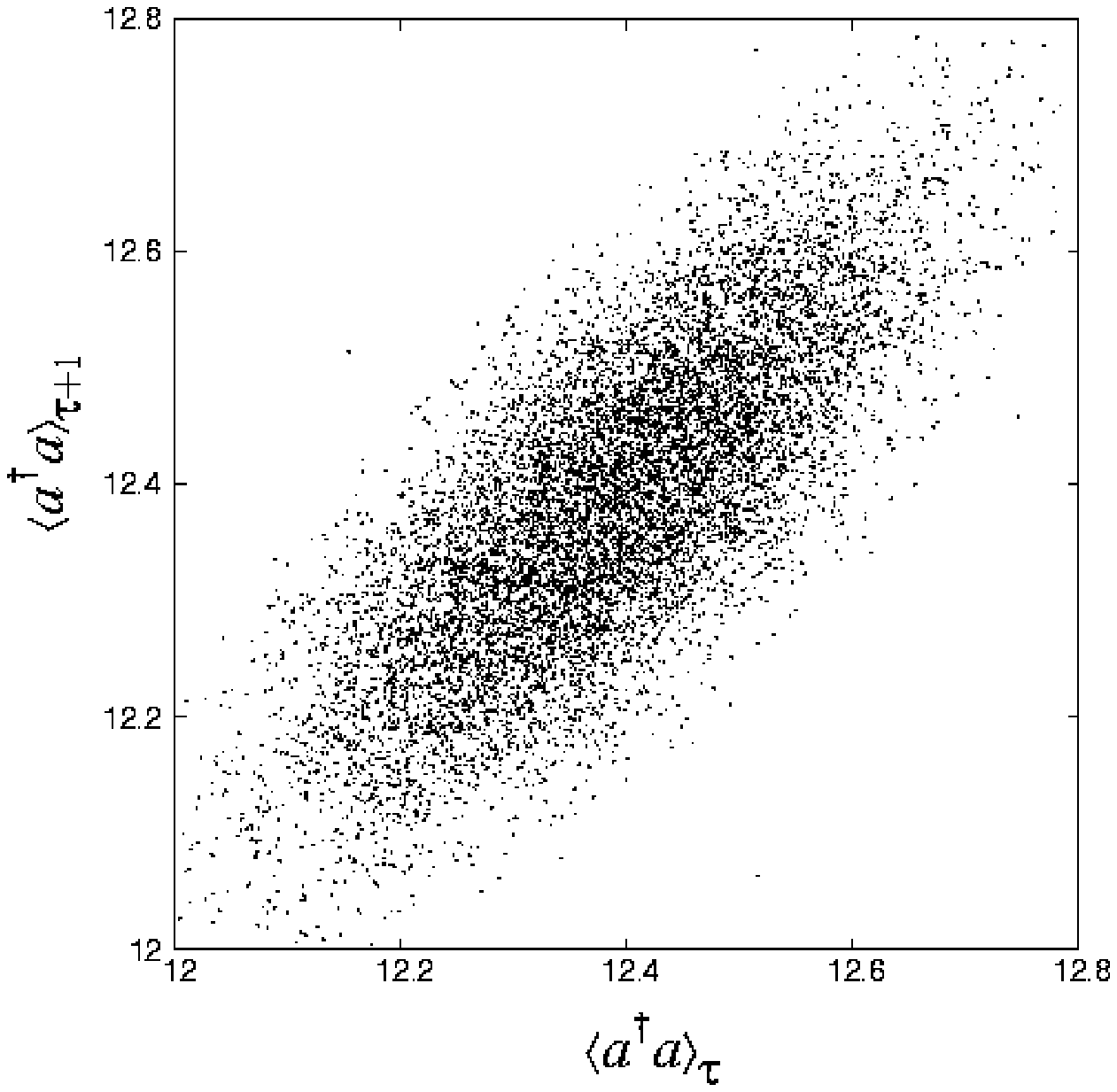}
\caption{\label{fig11} Return map of the mean photon 
number $\aver{\ad a}$ for the initial state  
$\ket{(\alpha,5);0}$ with $\nu = 5$; $\gamma/g = 5$.}
\end{figure}
\begin{figure}
\includegraphics[width=2.5in]{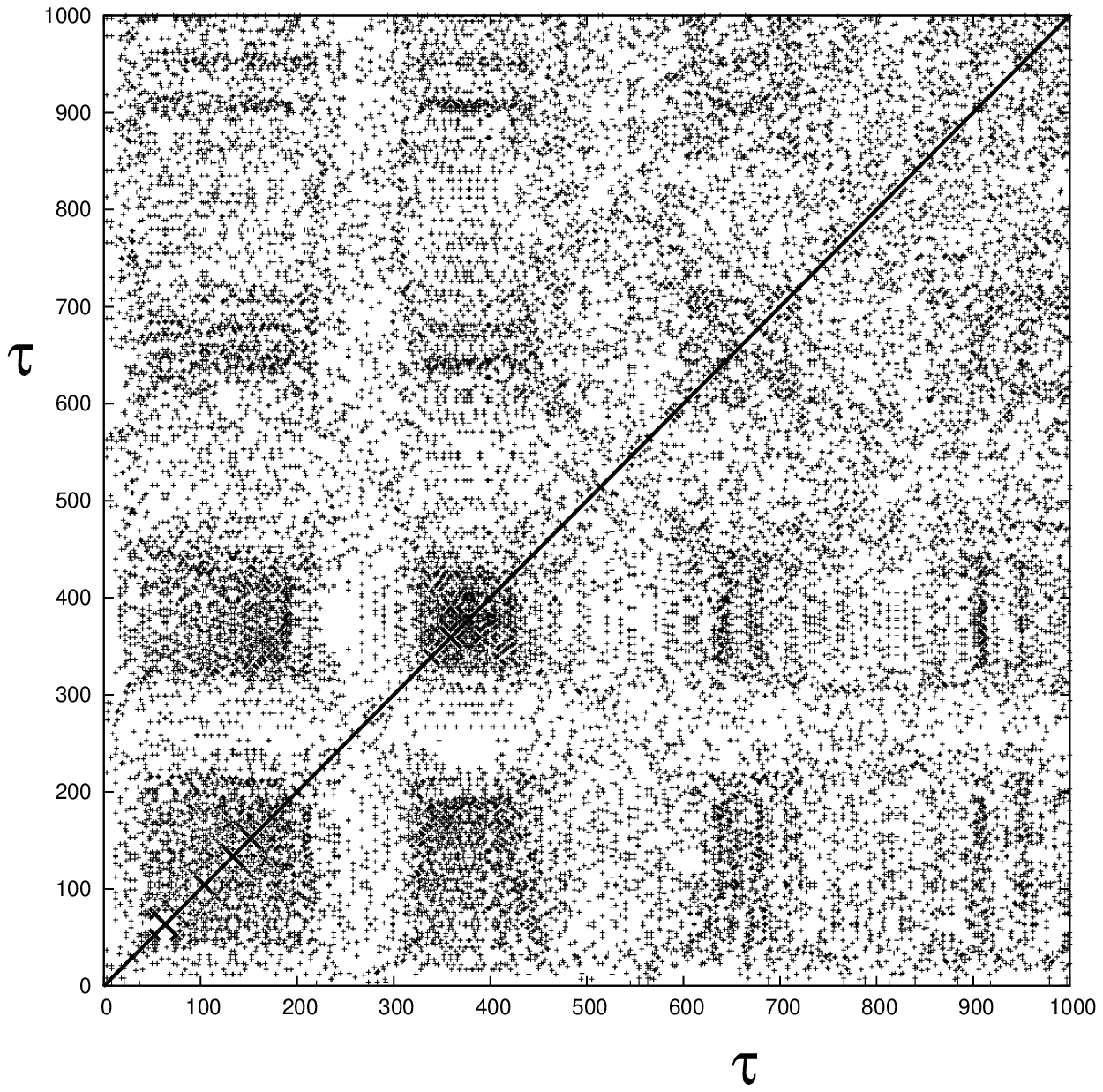}
\caption{\label{fig12} 
Recurrence plot of $\aver{\ad a}$ for 
the initial state $\ket{(\alpha,5);0}$ with $\nu = 5$; $\gamma/g = 5$.}
\end{figure}
The return map (Fig. \ref{fig11}) and the recurrence plot 
(Fig. \ref{fig12}) no longer have well-defined 
patterns. The first return time distribution to a 
cell $C = [12.455, 12.465]$, shown in   
Fig. \ref{fig13}, approaches an 
exponential distribution (dotted lines), while  
the invariant density of $\aver{\ad a}$,  depicted in 
Fig. \ref{fig14}, is consistent with a continuous 
distribution. 
\begin{figure}
\includegraphics[width=2.5in]{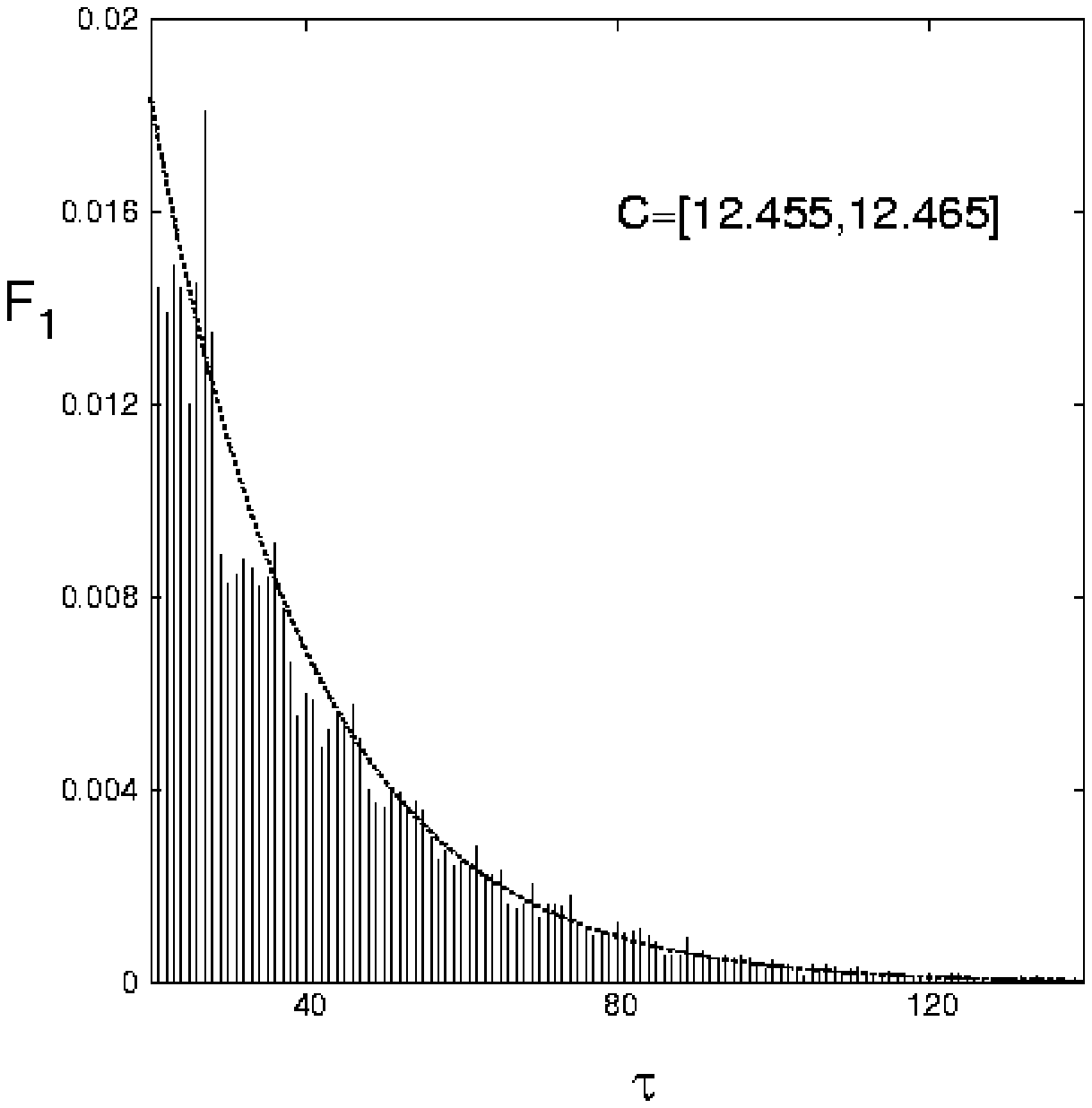}
\caption{\label{fig13} 
First return distribution $F_1$ of $\aver{\ad a}$ 
for the initial state 
$\ket{(\alpha,5);0}$ with $\nu = 5$; $\gamma/g = 5$.}
\end{figure}
\begin{figure}
\includegraphics[width=2.5in]{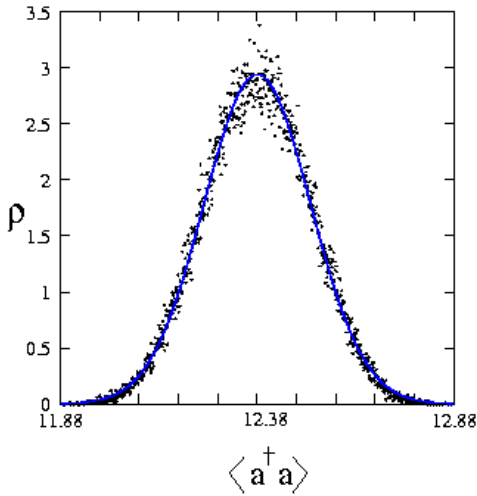}
\caption{\label{fig14} Invariant density: Initial state 
$\ket{(\alpha,5);0}$ with  $\nu = 5$; $\gamma/g = 5$.}
\end{figure}

As in the case of the single-mode model of Sec. \ref{singlemodemodel}, 
the analysis of recurrence time distributions and 
recurrence plots is supplemented by 
a time series analysis leading to the estimation 
of the minimum embedding dimension and the calculation of the
maximal Lyapunov exponent $\lambda_{\rm max}$.   
We have used\cite{sudh5} a robust algorithm developed
by Rosenstein {\it et al.}\cite{rosen} and
Kantz\cite{kantz}
for the reliable estimation of 
 $\lambda_{\rm max}$ from
data sets represented by time series.
Table \ref{table1} gives a 
capsule summary of the 
representative results on the two-mode model 
of this Section. 
\begin{table}
\includegraphics[width=3.5in] {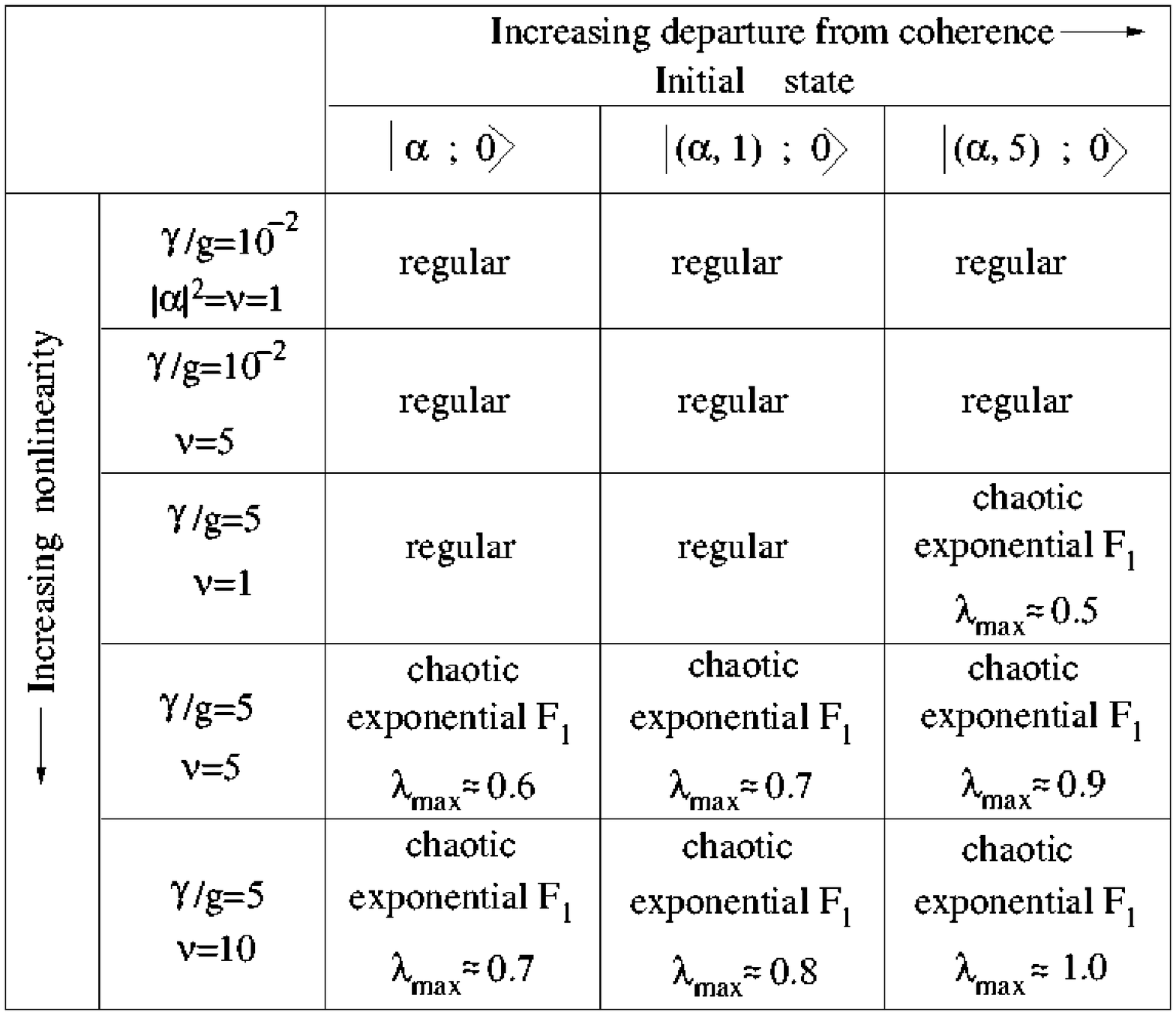}
\caption{\label{table1}  Qualitative dynamical behavior of the 
mean photon number of a 
single-mode electromagnetic 
field interacting with a nonlinear medium.} 
\end{table}
The characterizer `regular'  is used in those cases in which   
$\lambda_{\rm max} = 0$ to within numerical accuracy, while 
`chaotic' indicates those corresponding to a positive value 
of $\lambda_{\rm max}$. It is immediately apparent that 
the latter is accompanied by an exponential recurrence-time 
distribution, in keeping with the known results for hyperbolic 
dynamical systems.  As further corroboration we have also 
checked, in all these cases,  that the distribution of two successive 
recurrences fits the next term in a Poisson distribution. 
The trends in the foregoing results 
are also in agreement  
with the results presented elsewhere\cite{sudh4} 
on the entanglement properties of the model, including, in particular, 
the behavior of the subsystem entropy corresponding to the 
field mode.  

It is interesting that, even in as simple a 
model as the one considered here, where one of the components 
of a bipartite system effectively behaves like an open 
quantum system, the dynamical behavior can display 
such a range of ergodicity properties. The implications for 
genuine open quantum systems are therefore noteworthy.

\end{document}